\documentclass[12pt,english,aps,prb,longbibliography]{revtex4-1}
\usepackage[T1]{fontenc}
\usepackage[latin9]{inputenc}
\usepackage{color}
\usepackage{array}
\usepackage{float}
\usepackage{multirow}
\usepackage{amstext}
\usepackage{graphicx}
\PassOptionsToPackage{normalem}{ulem}
\usepackage{ulem}

\makeatletter

\providecommand{\tabularnewline}{\\}

\setcitestyle{super}

\usepackage[colorlinks=true,citecolor=blue,urlcolor=blue]{hyperref}

\usepackage{babel}
\usepackage{hyperref}

\makeatother

\usepackage{babel}
\begin{document}

\title{From Half-metal to Semiconductor: Electron-correlation Effects in
Zigzag SiC Nanoribbons From First Principles}

\author{Naresh Alaal$^{1,2,3}$, Vaideesh Loganathan $^{2}$, Nikhil Medhekar
$^{3}$, Alok Shukla $^{2,4,*}$}

\address{$^{1}$IITB-Monash Research Academy, Indian Institute of Technology
Bombay, Powai, Mumbai 400076, India}

\address{$^{2}$Department of Physics, Indian Institute of Technology Bombay,
Mumbai 400076, India}

\address{$^{3}$Department of Materials Engineering, Monash University, Clayton,
Victoria 3800, Australia}

\address{$^{4}$Present Address: Physics Department, Bennett University, Plot
No. 8-11, Tech. Zone II, Greater Noida 201310 (UP) India }
\email{shukla@phy.iitb.ac.in}

\begin{abstract}
We performed electronic structure calculations based on the first-principles
many-body theory approach in order to study quasiparticle band gaps,
and optical absorption spectra of hydrogen-passivated zigzag SiC nanoribbons.
Self-energy corrections are included using the GW approximation, and
excitonic effects are included using the Bethe-Salpeter equation.
We have systematically studied nanoribbons that have widths between
0.6 $\text{nm}$ and 2.2 $\text{nm}$. Quasiparticle corrections widened
the Kohn-Sham band gaps because of enhanced interaction effects, caused
by reduced dimensionality. Zigzag SiC nanoribbons with widths larger
than 1 nm, exhibit half-metallicity at the mean-field level. The self-energy
corrections increased band gaps substantially, thereby transforming
the half-metallic zigzag SiC nanoribbons, to narrow gap spin-polarized
semiconductors. Optical absorption spectra of these nanoribbons get
dramatically modified upon inclusion of electron-hole interactions,
and the narrowest ribbon exhibits strongly bound excitons, with binding
energy of 2.1 eV. \textcolor{black}{Thus, the narrowest zigzag SiC
nanoribbon has the potential to be used in optoelectronic devices
operating in the IR region of the spectrum, while the broader ones,
exhibiting spin polarization, can be utilized in spintronic applications. }

\end{abstract}
\maketitle

\section{{\normalsize{}Introduction}}

After the synthesis of graphene \cite{graphene_synth}, and the discovery
of its unique electronic, optical and thermal properties, research
on low-dimensional materials has increased tremendously. Graphene
is attractive because of its unique properties such as its high electronic
conductivity, and high mobility at room temperature \cite{gn1,gn2,gn3}.
However its applications are limited in semiconducting devices because
of its zero band gap. This gaplessness of graphene has motivated the
discovery of alternate nanomaterials that possess finite band gap,
and have the potential to replace the silicon in semiconductor technology.
Recently two-dimensional (2D) materials which have finite band gap
such as monolayers of hexagonal boron-nitride (h-BN) \cite{bn_synth},
transition metal chalcogenides (MoS$_{2}$, WS$_{2}$ etc.)\cite{mos2,opt_tmdc},
and phosphorene have been successfully synthesized\cite{phospho}.

Quasi one-dimensional materials such as nanoribbons, nanotubes, nanorods,
and nanowires have also attracted attention because of their interesting
photochemical, photophysical, and transport properties\cite{nr1,nr2}.
Nanoribbons (NRs), in particular, have received great attention from
researchers because of their unique electronic properties which can
be modified on the basis of their edge configuration, and width. Quantum
confinement because of the presence of edges, and their different
possible passivations, lead to interesting optical, electronic and
magnetic properties. Graphene nanoribbons (GNRs) and boron-nitride
nanoribbons (BNNRs) have been experimentally synthesized by unwrapping
of carbon nanotubes, and BN nanotubes, respectively\cite{gnr_synthl,bnnr_synth}.
Armchair GNRs (AGNRs) are non magnetic semiconductors for all widths,
and have oscillating band gaps over families, approaching the zero
band gap of a 2D sheet, for large widths\cite{AGNR_Eg_osc}. Zigzag
GNRs (ZGNRs), on the other hand, have tunable band gaps ranging from
metal to semiconductor, depending on the width, and passivation edges\cite{O_zgnr,pass_zgnr}.
Furthermore, ZGNRs also exhibit half-metallic behavior with spin-polarized
band gaps when an electric field is applied along the direction of
the width of the ribbons, and their gaps can be controlled by the
strength of the electric field\cite{ZGNR_Efield}.

\textcolor{black}{C and Si have same number of electrons, therefore
SiC structures are stable in bulk and have electronic properties which
are intermediate in between bulk Si and carbon structures. Silicon
carbide (SiC) crystallizes in several forms such as hexagonal, rhombohedral,
and cubic Bravais lattices\cite{edge_H_zig}. Bulk SiC is a wide band
gap semiconductor, and has been used in high-temperature, high-pressure
and high-frequency device applications\cite{SiC_app,SiC_Eg}. The
graphene-like 2D SiC monolayer has not been experimentally fabricated
yet, but has been extensively studied using theoretical methods}\cite{SiC_mono,ASiCNR_Eg_osc,sic_bare,sic2d_eg,SiX,sic_bulk_gw,SiC_BN_BeO,group4,opt_SiCNT,tight_bind}\textcolor{black}{.
Unlike graphene, it is a direct band gap semiconductor with a band
gap of 2.5 eV. An ultra-thin SiC nanosheet which has a thickness of
$0.5-1.5\text{ nm}$ has been fabricated, and is used in light emitting
applications }\cite{SiC2D}\textcolor{black}{. Various SiC structures
such as SiC nanotubes, SiC nanowires, microribbons, and crystalline
and bicrystalline nanobelts have also been successfully synthesized\cite{SiCNT_synth,sicnr_micro,3C_nanobelts,bicrystaline_nanobelt}.
Furthermore multilayer SiC nanoribbons in lengths of micrometers,
and thicknesses of nanometers have also been synthesized}\cite{SiCNR_synth}\textcolor{black}{. }

As far as theory is concerned\textcolor{black}{,{} SiC nanoribbons
have been extensively studied using }first-principles calculations,
based on the density functional theory approach (DFT)\textcolor{black}{{}
}\cite{ASiCNR_Eg_osc,ZSiCNR_DFT,ZSiCNR_Efield,assymetry_H,edge_H_zig,sic_bare,O_S_zig,mod_eg,B_N_SiCNR,half_bare_zig,vac_SiCNR}\textcolor{black}{.
}Sun et al. \cite{ASiCNR_Eg_osc} \textcolor{black}{reported that
hydrogen- passivated armchair SiCNRs (ASiCNRs) are direct band gap
semiconductors, with band gaps in the range of 2.3 - 2.4 eV. Furthermore,}
\textcolor{black}{they reported that band gaps of zigzag SiCNRs (ZSiCNRs)
are spin polarized, and exhibit intrinsic half-metallic behavior,
without the application of an external electric field.}\cite{ASiCNR_Eg_osc}
Lou \emph{et al}.\cite{ZSiCNR_DFT} studied narrow ZSiCNRs with widths
in the range 0.6\textendash 1.6 nm, and showed that the ferrimagnetic
state is more stable,\textcolor{cyan}{{} }with an anti parallel spin
orientation between the two edges. ZSiCNRs that have a half-bare-edge
were studied by Tang \emph{et al.}\cite{half_bare_zig} using hybrid
DFT approach. They reported that half-bare-edge ZSiCNRs with a bare
carbon edge are magnetic semiconductors, while the ones with bare
Si edge atoms are magnetic metals. The influence of doping with B
and N impurities has been studied by Costa \emph{et al.},\cite{B_N_SiCNR}
who showed that B doped ZSiCNRs retain half metallic behaviour, while
N doped ZSiCNRs become metallic. Lopez-Benzanilla\emph{ et} \emph{al.}\cite{O_S_zig}
employed a local spin-density approximation (LSDA) approach to study
sulfur and oxygen passivated ZSiCNRs to discover that, ZSiCNRs turned
from half-metallic to semiconductors, or metals, as a result of passivation.
The effects of substitution of edge atoms, with B and N atoms, has
been studied by Zheng \emph{et al.}\cite{mod_eg} who reported that
modified ZSiCNRs are semiconductors. Bekaroglu \emph{et al.,}\cite{sic_bare}
and Morbec \emph{et al.},\cite{vac_SiCNR} studied the effects of
vacancies on the electronic structure of ZSiCNRs by using the DFT-GGA
approach, and found that double (Si and C) vacancies induce magnetism.\textcolor{black}{{}
Ultra narrow ZSiCNRs were studied by Ping et al.\cite{lou_short}
including short-range exact exchange effects, and they obtained lower
edge energies, and higher band gaps, as compared to the ones computed
using GGA. }

\textcolor{black}{It is well known that the DFT-based approaches underestimate
the band gap in semiconductors, because they do not include electron
correlation effects.\cite{gw_acc} It has been demonstrated that in
nanoribbons, not including correlation effects causes severe errors
as compared to their bulk counterpart, because in low-dimensional
systems these effects are enhanced \cite{GNR_GW,GNR_GW2,GNR_GW3,GNR_GW4,GNR_GW5,AMoS2,Nalaal}.
It will be also be interesting to study many-body electron-electron
interactions in evaluating accurate band gaps and electron-hole effects
in order to calculate optical absorption spectra in ZSiCNRs; this
knowledge will enable the exploration of their potential in optical
device applications. In the present work, we calculate the quasiparticle
band structures and the optical absorption spectra of ZSiCNRs by using
GW approximation, and by solving the Bethe-Salpeter equation (BSE),
respectively. We find that ultra-narrow ZSiCNRs, with widths narrower
than 1 nm, are non-magnetic semiconductors. GW corrections open up
significant band gaps in ZSiCNRs that are wider than 1 nm, and transform
them from half-metals to semiconductors. We find that narrowest ZSiCNR,
with the width of 0.6 nm, has strongly bound excitons, with a binding
energy of 2.1 eV, while other nanoribbons have weakly bound excitons
because their band gaps are too small. Therefore, the narrowest ribbon
can be utilized in optoelectronic applications in the infrared frequency
range. Broader nanoribbons exhibit distinct band gaps for the two
spin orientations, and, therefore, they can be utilized in spintronic
devices such as spin valves.}

\textcolor{black}{The remainder of this paper is organized as follows.
Section \ref{sec:theory} presents details of computations employed
in this work, while section \ref{sec:results} presents results of
our calculations on ZSiCNRs of varying widths. Finally, in section
\ref{sec:conclusions}, we present our conclusions.}

\section{{\normalsize{}Computational Details}}

\label{sec:theory}

\textcolor{black}{We carried out first-principles many-body electronic
structure calculations for ZSiCNRs using a three-step procedure. Firstly,
the DFT calculations are performed using a plane wave approach, in
the generalized gradient approximation (GGA),\cite{gga} with Perdew-Burke-Enzerhof
(PBE) pseudopotentials.\cite{psp} Geometry optimizations were carried
out using the computer program Vienna ab initio simulation package
(VASP) \cite{VASP}. All the nanoribbons considered here were fully
relaxed until the force on each atom was lower than 0.01 eV/\AA.
The threshold energy convergence was set to $10^{-4}$ eV. A Monokhorst
k-point grid of $1\times1\times11$ was used for Brillouin zone integration
during the relaxation. After the relaxation, we used optimized atomic
positions and lattice parameters to calculate Kohn-Sham band gaps
of ZSiCNRs using the DFT-GGA approach as implemented in the software
package ABINIT \cite{abinit1,abinit2}. GGA wave functions are a good
starting point for the many-body calculations, aimed at including
electron correlation effetcs.}

\textcolor{black}{In the second step we calculate quasiparticle corrections
for the DFT-GGA band gaps using the many-body approach G$_{0}$W$_{0}$
approximation\cite{g0w0}, which is a single-shot GW approach for
computing self-energy corrections in terms of the Green's function
(G), and the screened Coulomb potential (W). The screening effects
were included by using the plasmon pole model\cite{ppm}. A kinetic
energy cutoff of 26 Ha has been used for all the systems considered
here. Cutoff energies of 5 Ha, and 12 Ha have been used for the correlation,
and the exchange, parts of self-energy calculations, respectively.
A total number of 200 bands were used, out of which more than 120
were unoccupied. In the last step, the BSE is solved within the Tamm-Dancoff
approximation\cite{tda} in order to obtain the optical response inclusive
of the contribution of the electron-hole interactions, retaining only
the resonant part of the Bethe-Salpeter Hamiltonian: This is because
the inclusion of the coupling has been shown to have a negligible
effect. This approximation has been also used to study similar systems
such as graphene nanoribbons,\cite{GNR_GW,GNR_GW6} and boron-nitride
nanoribbons\cite{GNR_GW4}. In the GW, and the BSE calculations, a
Coulomb truncation scheme was also used to avoid the inclusion of
long range interactions between the periodic images. A $1\times1\times30$
k-point grid has been used in the the GW and BSE calculations for
the systems considered here. Five valence band and five conduction
bands were included to obtain the optical absorption spectra. Nanoribbons
were considered to be periodic along the $z$ direction, and more
than 13 \AA\ wide vacuum layers were taken in non-periodic directions
($x$ and $y$ directions), in order to represent an isolated system.
Width of a given ZSiCNR was specified by prefixing it with the number
of zigzag lines ($N_{z}$) across its width. Thus ZSiCNR-3 denotes
a ribbon with $N_{z}$=3 zigzag lines across the width. In the present
work, we have studied ribbons with $2\leq N_{z}\leq8$, which corresponds
to widths between 0.6 $\text{nm}$ and 2.2 $\text{nm}$.}

\section{{\normalsize{}Results And Discussion}}

\label{sec:results}

\subsection{Formation energy and stability}

\begin{figure}
\begin{centering}
\includegraphics{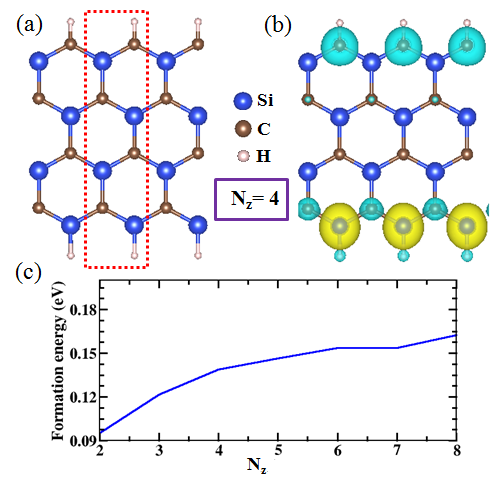}
\par\end{centering}
\caption{(a) Geometric structure of ZSiCNR-4 unit cell shown in the red dashed
box is considered for the calculation\textcolor{red}{. }(b) Calculated
spin-density of ZSiCNR-4. Blue and yellow colored spin densities denote
opposite spin orientations. (c) Computed edge formation energy of
ZSiCNRs as a function of width. }

\label{str_zsicnr}
\end{figure}

Fig. \ref{str_zsicnr} (a) shows the geometric structure of ZSiCNR-4.
Optimized bond lengths corresponding to Si-C, Si-H, and C-H were obtained
to be 1.77 \AA, 1.49 \AA, and 1.09 \AA, respectively, while the
lattice constant was found to be 3.11 \AA, for all the structures
considered here. These parameters are in good agreement with previous
studies\cite{ASiCNR_Eg_osc,lou_short}. Here, we calculate the edge
formation energies of these nanoribbons in order to understand their
thermodynamic stability, during edge formation. We evaluated the formation
energy by using the following formula:

\textcolor{black}{
\[
E_{f}=\left(E_{T}-N_{SiC}E_{SiC}-0.5N_{H}E_{H_{2}}\right)/2l,
\]
where $E_{T}$ is the total energy of the hydrogen passivated zigzag
SiC nanoribbon , $E_{SiC}$ is the total energy of the 2D SiC sheet,
$E_{H_{2}}$ is the total energy of an isolated hydrogen molecule,
$N_{SiC}$ is the number of SiC dimers in the unit cell, $N_{H}$
is the number of H atoms in the unit cell, and $l$ is the periodic
length of an edge. In general, the lower the value of $E_{f}$ of
a material, the higher its stability}. The formation energy of ZSiCNRs
as a function of width has been plotted in Fig. \ref{str_zsicnr}
(c). Unlike ASiCNRs,\cite{Nalaal} formation energy of these nanoribbons
is strongly dependent on their width, and it increases with the width
of the nanoribbons. From the figure it is obvious that ribbons which
have narrow widths have high stability, suggesting that they will
be relatively easier to synthesize in the laboratory.

\subsection{Nonmagnetic ZSiCNRs}

\subsubsection{Quasiparticle energies}

\begin{figure}[H]
\begin{centering}
\includegraphics[scale=1.2]{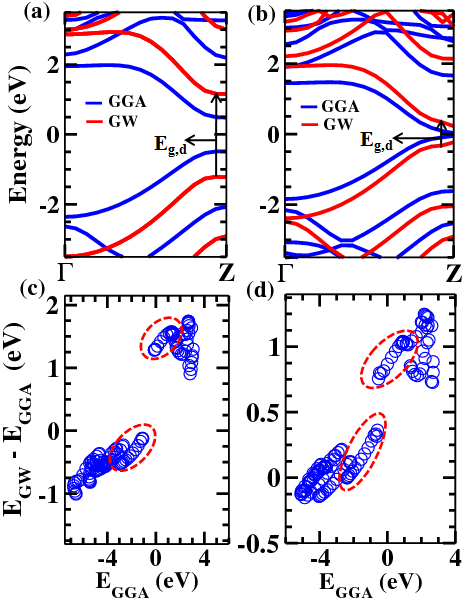}
\par\end{centering}
\caption{(a) and (b) GGA and GW band structures of ZSiCNR-2 and ZSiCNR-3. Blue
and red lines represent GGA and GW band structures, respectively,
and black arrows denote interband transitions. (c) and (d) Quasiparticle
self-energy corrections (differences between quasiparticle energies
and GGA energy values) for various bands to the GGA Kohn-Sham energies
for ZSiCNR-2, and ZSiCNR-3, respectively. Corrections to the states
corresponding to the edge atoms in valence, and conduction, bands
are enclosed in the red dashed ellipses.}

\label{band_z2z3}
\end{figure}

First we discuss the electronic structure of ZSiCNRs with widths $N_{z}=\text{2}$
and $\text{3}$, which do not exhibit any magnetic behavior. Ping
\emph{et al.} \cite{lou_short} reported that these two nanoribbons
are non-magnetic semiconductors, which do not exhibit half-metallic
behaviour within the GGA or HSE approaches, unlike ribbons which are
wider than 1 nm. Their DFT-GGA study predicted that ZSiCNR-2 is a
non-magnetic semiconductor with a band gap of 0.97 eV, and ZSiCNR-3
is essentially metallic with a band gap of 0.07 eV. Our calculated
DFT-GGA and GW band structures of ZSiCNR-2 and ZSiCNR-3 are shown
in Figs. \ref{band_z2z3} (a) and (b ) respectively. The blue lines
correspond to the DFT-GGA, while the red lines denote the GW band
structure. Our DFT results agree with those of Ping et al \cite{lou_short}.
Fig. \ref{band_z2z3} (a) shows that ZSiCNR-2 is a direct band gap
material at the DFT-GGA level, and the gap occurs at the $Z$ point
located at the edge of Brillioun zone. After including the many-body
effects, computed band structure is depicted by the red lines. We
found that quasiparticle corrections increase the band gap from 0.97
to 2.4 eV, which is an indirect one (quasi-direct) at the GW level.
In the GW band structure of ZSiCNR-2, the valence band maximum (VBM)
occurs small distance away from the $Z$ point, and there is no change
in the conduction band minimum (CBM). Fig. \ref{band_z2z3} (b) shows
the band structures computed by GGA and GW approaches, for ZSiCNR-3.
Self-energy corrections widened the band gap from 0.07 eV to 0.45
eV, which is a six fold increase due to electron correlation effects.
This increase shows that many-body corrections transform ZSiCNR-3
from a nearly metallic system, to a semiconductor.

Self-energy corrections for various bands for both the ribbons are
shown in Figs. \ref{band_z2z3} (c) and (d). In these nanoribbons,
$\pi$ band states corresponding to the $p_{x}$ orbital extend into
the vacuum perpendicular to the ribbon plane, while the $\sigma$
states are composed of $p_{z}$ and $p_{y}$ orbitals lying in the
plane of the sheet. The self-energy corrections to the $\pi$ states
are larger than those to the $\sigma$ states because of the screening
effects due to other $\pi$ electrons . The corrections enclosed in
the red ellipses correspond to the edge states which experience enhanced
Coulombic interactions. When compared to GGA band structure, GW bands
not only shift but also stretch by an average of 20\%-30\% because
of non-uniform quasiparticle corrections. \textcolor{black}{We find
that corrections to the states which derive main contribution from
the interior atoms are less linear when compared to states which are
predominantly localized on edge atoms}\textcolor{black}{\uline{.}}

\begin{figure}[H]
\begin{centering}
\includegraphics[scale=0.6]{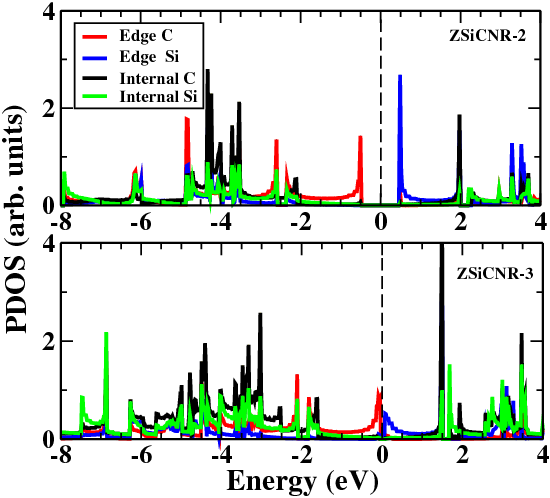}
\par\end{centering}
\caption{\textcolor{black}{Projected density of states for ZSiCNR-2 and ZSiCNR-3.
Red and blue curves show states which derive contributions from the
edge C and Si atoms, while black and green curves represent states
which derive main contributions from internal C and Si atoms, respectively.
Fermi level is represented by black dashed line, which has been set
to 0 eV.}}

\label{fig:pdos_z2_z3}
\end{figure}
 \textcolor{black}{We present projected density of states (PDOS) for
ZSiCNR-2 and ZSiCNR-3 in Fig. \ref{fig:pdos_z2_z3}, in order to illustrate
the contribution of various atoms from the edges, and the interior,
to different orbitals. PDOS plots show that the VBM in both the ribbons
arises from edge carbon atoms, while the CBM is derived from edge
Si atoms. The states located on internal C and Si atoms are away from
the Fermi-level in both the ribbons.}


\subsubsection{Optical absorption spectra}

\begin{figure}[H]
\begin{centering}
\includegraphics[scale=0.5]{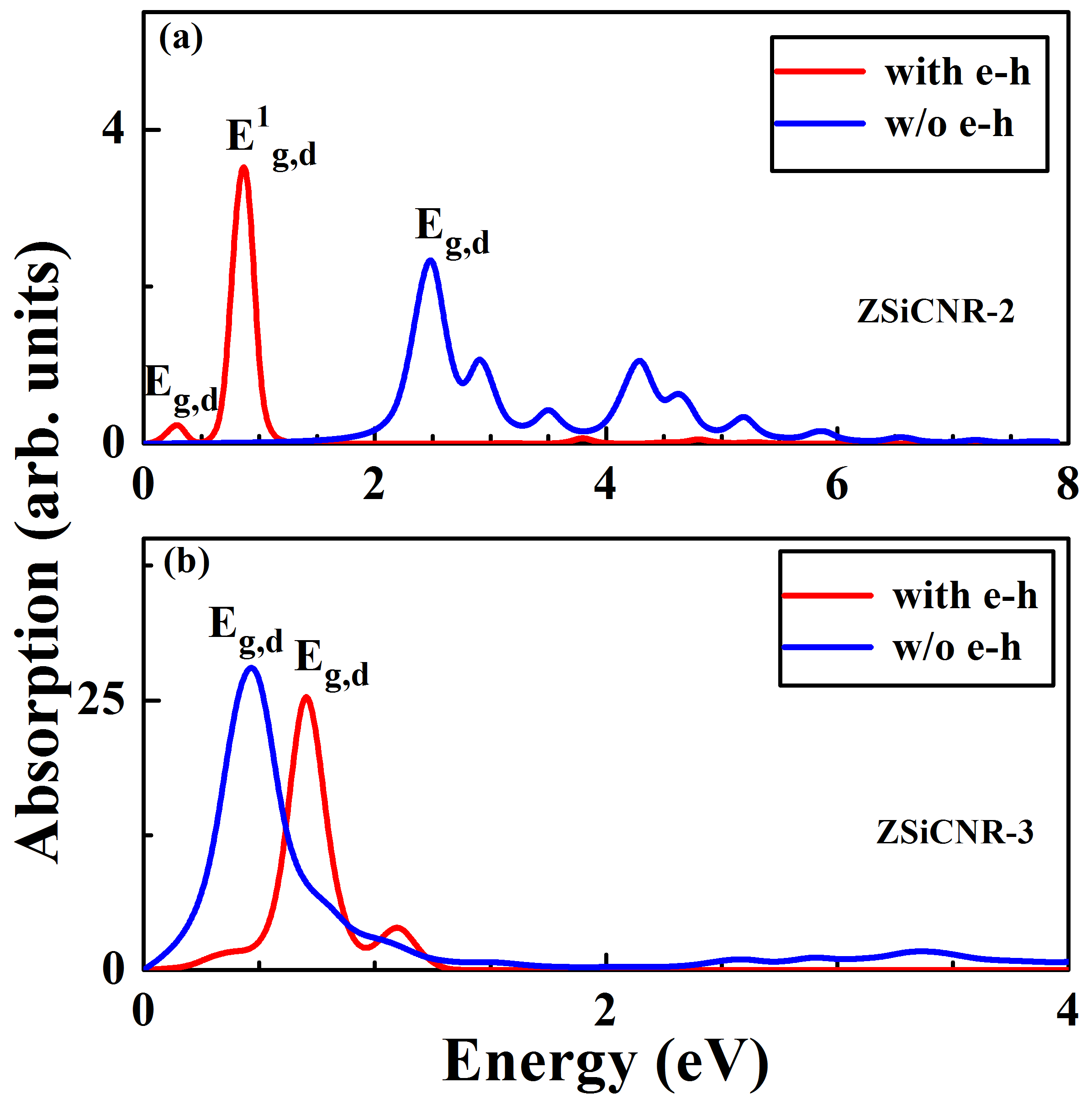}
\par\end{centering}
\caption{\textcolor{black}{Optical absorption spectra of (a) ZSiCNR-2, (b)
ZSiCNR-3. The blue curve shows the spectrum computed without the contribution
of electron-hole interactions, while the red curve represents spectrum
calculated by including electron-hole interactions.}}
\centering{}\label{fig:bse_zsicnr}
\end{figure}

\textcolor{black}{Using the quasiparticle band structure, we calculated
the optical absorption spectrum of the ZSiCNR-2 and ZSiCNR-3, with,
and without electron-hole interactions as is shown in Figs. \ref{fig:bse_zsicnr}
(a) and (b). Electron-hole interactions were incorporated by using
the BSE, and in all the calculations the polarization direction of
incident light was considered to be along the length of the nanoribbons.
We use $E_{g,d}$ notation to label the direct transitions which occur
between first valence band and first conduction band as shown in Fig.
\ref{band_z2z3}. The blue curve represents absorption due to interband
transitions at the GW-RPA level, while the red curve is a result of
solving the BSE, including electron-hole effects. In both the ribbons,
the first prominent peak comes from the transition between the first
valence band, and the first conduction band. In GW-RPA absorption
spectra we can observe that there are several peaks due to different
interband transitions in ZSiCNR-2. However, there is only one prominent
peak in ZSiCNR-3 which is a result of the transition between the VBM
and the CBM. The inclusion of electron-hole interactions changes the
spectrum significantly because, not only the positions of the peaks
are shifted, but also their shape looks different, when compared with
the GW-RPA absorption spectrum. In the absorption spectra of both
the ribbons, which include electron-hole interactions, we observe
only one intense excitonic peak due to the transition between first
valence and first conduction band, $E_{g,d}$. We did not find any
other transitions among other bands, even though we included five
valence and five conduction bands in our calculations. In Fig. \ref{fig:bse_zsicnr}
it is obvious that ZSiCNR-2 has a strongly bound exciton, of binding
energy of 2.1 eV ($E_{g,d}$), but with relatively weaker intensity,
corresponding to the first absorption peak. We also observed another
strongly bound excitonic peak ($E_{g,d}^{1}$), with a very high intensity,
and binding energy of 1.5 eV, originating at distinct $k$-points
as compared to the first one. These binding energies are significantly
larger than the calculated exciton binding energy, 1.17 eV, for the
2D SiC sheet,\cite{sic_bulk_gw} suggesting that the reduced dimensionality,
and quantum confinement, lead to an enhancement in the exciton binding
energies of nanoribbons. From among all the ribbons considered here,
it is clear that ZSiCNR-2 is a small gap semiconductor which can be
used in optical applications in the low energy region of spectrum.
Unlike ZSiCNR-2, strongly bound excitons are not observed in ZSiCNR-3
because the band gap is much smaller for this ribbon. Band to band
absorption spectrum peak at GW level gets blue shifted upon inclusion
of electron-hole effects, as shown in Fig. 4. The final location of
the blueshifted absorption peak clearly implies negative exciton binding
energy. Thus, the nature of optical transitions contributing to various
peaks is also modified in ZSiCNR-3, once the many-body effects are
included.}

\subsection{Spin-polarized ZSiCNRs}

\begin{figure}[H]
\begin{centering}
\includegraphics[scale=0.5]{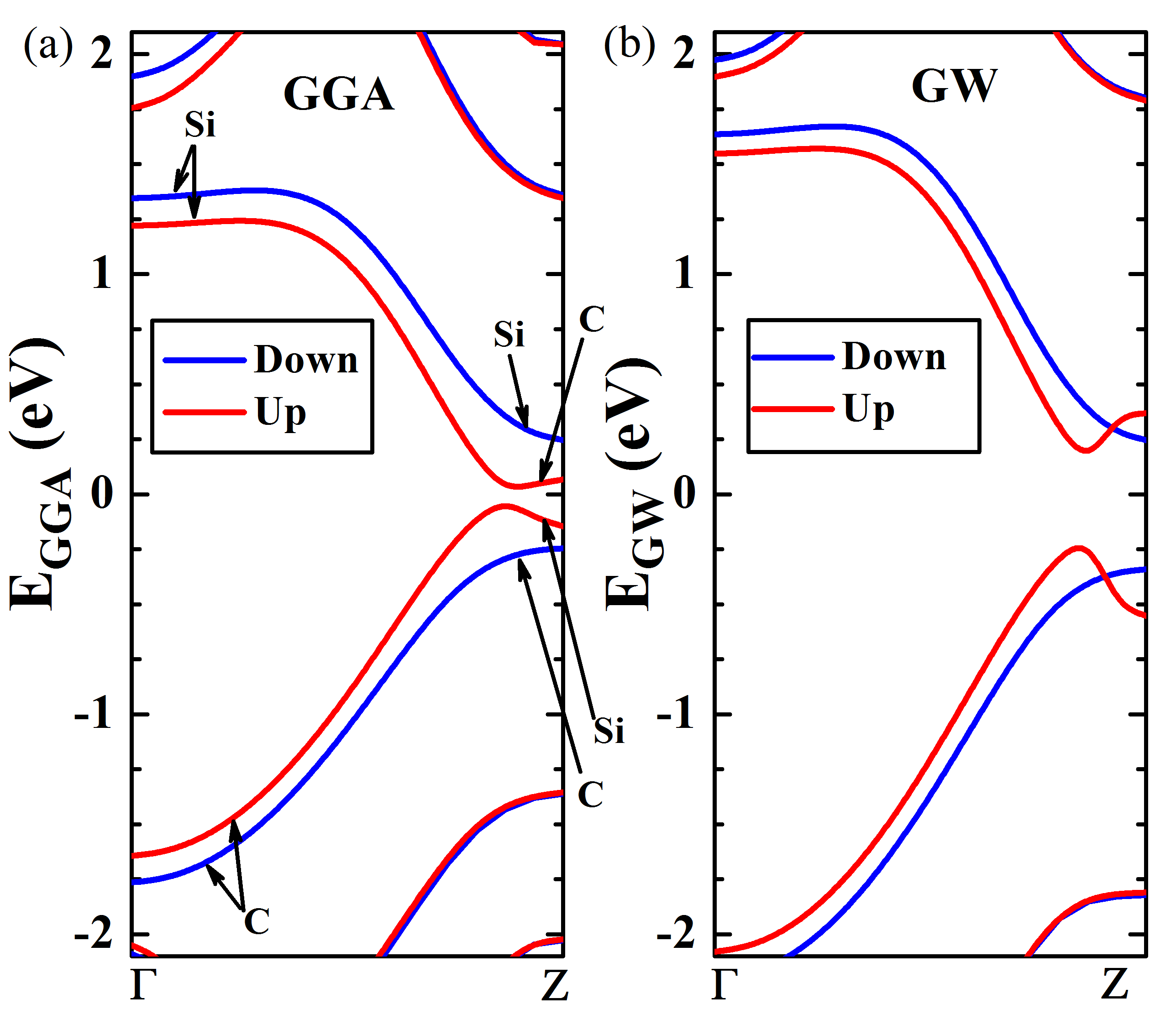}
\par\end{centering}
\caption{\textcolor{black}{(a) GGA and (b) quasiparticle band structures of
ZSiCNR-4. Red and blue lines represent spin-up and spin-down states,
respectively. The atoms contributing to the states of the valence
and conduction band on either side of the band gap are marked for
ZSiCNR-4. Towards the zone boundary, the states are localized at the
ribbon edges.}}

\label{band_znr4}
\end{figure}

\begin{figure}[H]
\begin{centering}
\includegraphics[scale=0.7]{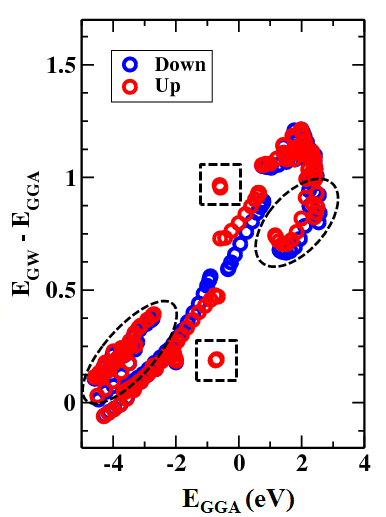}
\par\end{centering}
\caption{\textcolor{black}{Quasiparticle self-energy corrections to the GGA
Kohn-Sham energies of ZSiCNR-4. Blue and red empty circles represent
spin-up, and spin-down, states respectively. Corrections to the spin-up
edge states are shown in the dashed rectangular boxes. The dashed
ellipses enclose the $\sigma$ states.}}

\label{fig:structure}
\end{figure}

\textcolor{black}{Next we present the results of quasiparticle band
structure calculations of spin-polarized ZSiCNR-4, because $N_{z}=4$
to $N_{z}=8$ have similar band structures, and at mean-field DFT
level exhibit intrinsic half-metallic behavior.\cite{ASiCNR_Eg_osc,ZSiCNR_DFT}
In Figs. \ref{band_znr4} (a) and (b) we compare the GGA and GW quasiparticle
band structures of ZSiCNR-4, and show the widened quasiparticle band
gap, when electron-correlation effects are included. While DFT-GGA
predicts a very small gap of 30 meV, the GW quasiparticle gap is seen
to have widened to 0.33 eV, clearly showing that the ZSiCNRs are narrow-gap
semiconductors (Table \ref{tab-gaps1}). Although, after including
many-body effects through GW calculations, half-metallicity disappears,
but still the band gaps for two spin orientations are different. The
atoms contributing to various states are marked in Fig. \ref{band_znr4}
(a). Towards the zone boundary, the states of the first valence and
conduction bands of both spins are localized at the ribbon edges.\cite{ZSiCNR_DFT}
From the spin-density plot presented in Fig. \ref{str_zsicnr} (b),
it is obvious that the spins localized on the two edges have opposite
spin orientations. The states to the left of the VBM/CBM extend over
the entire ribbon width, except for the spin-down conduction band
which mainly comprises of the Si edge states. The screening at the
edges shows a complicated dependence on the states, which is enhanced
in ZSiCNRs because the charge and spin distributions differ at the
Si and C edges. Fig. \ref{fig:structure} shows that the self energy
corrections are highly state-dependent. The states which are enclosed
in rectangular boxes, away from the main region, represent the spin-up
edge states (valence states at the Si edge, and conduction states
at the C edge). As the wave vector approaches the zone boundary (Z
point), the states get more localized at the edges,\cite{ZNR_edge_states}
and the deviations between up- and down-spin states, as also the energy
gap for up-spin states, become larger (}\textcolor{black}{\emph{cf.}}\textcolor{black}{{}
Fig. \ref{band_znr4}(a)). Thus, the nature of spin-up bands is very
similar to the bands in spin degenerate zigzag graphene} nanoribbons.\cite{GNR_GW3}

\begin{table}
\begin{tabular}{ccccccc}
\hline 
 &  & spin-up  &  &  & spin-down  & \tabularnewline
\hline 
\textcolor{black}{N$_{z}$ } & $E_{GGA}^{gap}$  & $E_{GW}^{gap}$  & $\Delta E^{gap}$  & $E_{2\ GGA}^{gap}$  & $E_{2\ GW}^{gap}$  & $\Delta E_{2}^{gap}$\tabularnewline
\hline 
4  & 0.03  & 0.33  & 0.30  & 0.55  & 0.59 & 0.04\tabularnewline
5  & 0.06  & 0.82  & 0.76  & 0.72  & 0.77 & 0.05\tabularnewline
6  & 0.006  & 0.38  & 0.37  & 0.64 & 0.74  & 0.1\tabularnewline
7 & 0.002 & 0.384 & 0.38 & 0.64 & 0.92 & 0.28\tabularnewline
8 & 0.08 & 0.77 & 0.69 & 0.76 & 0.93 & 0.14\tabularnewline
\hline 
\end{tabular}

\caption{\textcolor{black}{The values of GGA and GW band gaps, and their difference,
for spin-polarized ZSiCNRs for widths N$_{z}$= 4 to 8. Columns 2-4
(Columns 5-7) correspond to spin-up (spin-down) channel. All energies
are in eV. \label{tab-gaps1}}}
\end{table}

\vspace{0.2in}


\vspace{0.2in}

The larger spin-down gap at the zone boundary exhibits a smaller quasi-particle
correction. Fig. \ref{fig:structure} shows the valence spin-down
states bending away from the spin-up states, and, similarly, spin-down
states of the conduction band too have smaller corrections than the
corresponding spin-up states. These states correspond to the valence
C, and the conduction Si, states localized at the edges. The spin-down
edge states show smaller self energy corrections when compared to
those localized in the interior, while the spin-up edge states exhibit
exactly the opposite behavior. The charge distributions are different
at the two edges, leading to varying amounts of screening. For example,
in the ground state, there is a partial valence electron transfer
from Si to C.\cite{SiCnr_polar} As a consequence, the edge C states
are screened more, and have smaller self energy corrections, while
the lesser screened edge valence Si states experience enhanced Coulomb
interactions. Therefore, the spin-down band gap does not change as
much as the spin-up band gap. 

A few points enclosed by dashed circles lying outside the main region
are $\sigma$ states. Apart from the first valence and conduction
bands, the next few bands comprise both $\sigma$ and $\pi$ states
(shown in Fig. \ref{fig:structure}). Similar to ASiCNRs,\cite{Nalaal}
and narrow ZSiCNRs, the $\sigma$ states have smaller-self energy
corrections than the $\pi$ states. 

\textcolor{black}{For the sake of completeness, we also performed
calculations on the non-magnetic (i.e., non-spin polarized) states
of ZSiCNRs, and found that spin polarized state does not exist for
narrow ribbons ($N_{z}=2,\:3$), and spin-polarized calculations for
those ribbons yield the same results as non-magnetic calculations.
However, for broader ribbons, both spin-polarized and non-magnetic
states exist, and quite expectedly, total energies for the spin-polarized
cases are lower as compared to the non-magnetic ones (see Table \ref{tab:mvsnm}).
We also performed GGA and GW calculations for the non-magnetic states
of broader ribbons and results are presented in Table \ref{tab:nonmag}.
From the table it is obvious that: (a) at DFT-GGA level band gaps
of the ribbons decrease rapidly with their width, and (b) quasiparticle
corrections are large for the ultra narrow ZSiCNRs, but they also
reduce in magnitude rapidly with the increasing width. Combining these
two results, we conclude that non-magnetic calculations predict that
broader ZSiCNRs will exhibit metallic behavior. Therefore, an experimental
measurement of the band gaps of broader ZSiCNRs can settle the issue
whether these systems exhibit a magnetic ground state, or a nonmagnetic
one.}

\begin{table}
\begin{tabular}{c|cc|c}
\hline 
\multirow{2}{*}{\textcolor{black}{N$_{z}$}} & \multicolumn{2}{c|}{\textcolor{black}{Total energy (eV) }} & \multirow{2}{*}{\textcolor{black}{$\Delta E$ (eV)}}\tabularnewline
\cline{2-3} 
 & \textcolor{black}{NM} & \textcolor{black}{SP} & \tabularnewline
\hline 
\textcolor{black}{4 } & \textcolor{black}{-62.106675} & \textcolor{black}{-62.115140} & \textcolor{black}{0.008465}\tabularnewline
\textcolor{black}{5 } & \textcolor{black}{-76.101117} & \textcolor{black}{-76.120842} & \textcolor{black}{0.019725}\tabularnewline
\textcolor{black}{6 } & \textcolor{black}{-90.119364} & \textcolor{black}{-90.144896} & \textcolor{black}{0.025532}\tabularnewline
\hline 
\end{tabular}

\caption{\textcolor{black}{Total energies per unit cell for non-magnetic (NM)
and spin-polarized (SP) ZSiCNRs for widths N$_{z}$= 4 to 6. Last
column gives energy difference between both the configurations}\textcolor{green}{.
\label{tab:mvsnm}}}
\end{table}

\begin{table}
\begin{tabular}{cccc}
\hline 
\textcolor{black}{N$_{z}$ } & $E_{GGA}^{gap}$  & $E_{GW}^{gap}$  & $\Delta E^{gap}$ \tabularnewline
\hline 
2 & 0.97 & 2.4 & 1.43\tabularnewline
3 & 0.07 & 0.45 & 0.36\tabularnewline
4  & 0.011 & 0.22 & 0.209\tabularnewline
5  & 0.015 & 0.15  & 0.135\tabularnewline
6  & 0.002 & 0.07  & 0.068\tabularnewline
\hline 
\end{tabular}

\caption{\textcolor{black}{The values of GGA and GW band gaps, and their difference,
for non-magnetic states of ZSiCNRs for widths N$_{z}$= 2 to 6 . All
energy values are in eV.}\textcolor{black}{\uline{\label{tab:nonmag}}}}
\end{table}

\textcolor{black}{We have not presented optical absorption spectra
of spin-polarized ZSiCNRs as they undergo different quasiparticle
self-energy corrections because of two spin band gaps. We could expect
weakly bound excitons in ribbons that are wider than 1 $\text{nm}$.
Small band gap of ZSiCNRs which are wider than 0.6 $\text{nm}$ makes
them potentially useful in IR devices.}

\section{Conclusions}

\label{sec:conclusions}

\textcolor{black}{We performed first-principles many-body calculations
in order to investigate quasi particle band structures and optical
absorption of hydrogen-passivated zigzag SiC nanoribbons whose widths
ranged from 0.6 nm to 2.2 nm. For the study, computationally intensive
GW approximation was employed to compute the self-energy corrected
quasiparticle band gaps. We also used the BSE to calculate optical
absorption spectra including eletron-hole effects. It was found that
the many-body effects are significant due to the reduced dimensionality
of nanoribbons. Self-energy corrections transformed nearly half-metallic
zigzag SiC nanoribbons with width larger than 1 $\text{nm}$, to semiconductors.
The inclusion of electron-hole effects changed optical absorption
spectra in a significant manner, both qualitatively, and quantitatively.
The excitonic binding energies and luminescence properties are dependent
on width of the ribbon. We found that nanoribbon with a width of 0.6
$\text{nm}$ has strongly bound excitons with binding energy of 2.1
eV, suggesting its possible utility in optoelectronic applications.
We also computed the edge formation energy and showed that narrower
ribbons are more stable when compared to wider ones. The unique Coulombic
interactions at the spin-polarized edges make zigzag SiC nanoribbons
interesting systems to study. As far as behavior of their band gaps
with respect to the width is concerned, because the band gaps in half-metallic
ZSiCNRs arise due to the presence of magnetic edge states, therefore,
broader ribbons are expected to have spin-polarized non-zero band
gaps, hinting at their possible utility in spintronic applications,
such as spin valves. }
\begin{acknowledgments}
NA and NM gratefully acknowledge the support from Monash HPC, National
Computing Infrastructure of Australia, and the Pawsey Supercomputing
facility. This research was partially supported by the Australian
Research Council Centre of Excellence in Future Low-Energy Electronics
Technologies (project number CE170100039) and funded by the Australian
Government. AS acknowledges the financial support from Department
of Science and Technology, Government of India, under project no.
SB/S2/CMP-066/2013.
\end{acknowledgments}

\bibliography{sicznr}
\newpage{}
\end{document}